# Labels discover physics: the development of new labelling methods as a promising research field for applied physics

**Amelia Sparavigna**
Dipartimento di Fisica, Politecnico di Torino
Corso Duca degli Abruzzi 24, Torino, Italy
e-mail: amelia.sparavigna@polito.it

**Abstract**
Labels and tags are accompanying us in almost each moment of our life and everywhere we are going, in the form of electronic keys or money, or simply as labels on products we are buying in shops and markets. The label diffusion, rapidly increasing for logistic reasons in the actual global market, carries huge amount of information but it is demanding security and anti-fraud systems. The first crucial point, for the consumer and producer safety, is to ensure the authenticity of the labelled products with systems against counterfeiting and piracy. Recent anti-fraud techniques are based on a sophisticated use of physical effects, from holograms till magnetic resonance or tunnel transitions between atomic sublevels. In this paper we will discuss labels and anti-fraud technologies as a new and very promising research field for applied physics.

**Introduction**
Today, it is possible to obtain a huge variety of products by the simple use of a code, for instance the credit card number, directly in shops or on the World Wide Web. This simple action has behind it the now global organisation of the supply chain. In this organisation, the role of consumers is rapidly changing from the previously simple passive position to an active participation, actually pulling goods to meet their needs. It is a buyer's market, where the consumer wants to have a guarantee on goods quality, origin and safety. On the supply chain, and in particular on its logistic, it is now focussing a lot of researches, where the new technologies can be applied giving advantages to producers, retailers, suppliers and consumers.
In a world of distributed and mobile computing, extending the information technology to each point of the supply chain from producer to consumer is the key to achieve a sustainable competitive advantage.

It is clear that, as more and more applications use Auto identification (AutoID) technology, the demand for increasingly complex capability will drive to new technologies or to combinations of existing technologies, where each contributes with its key strengths. The joint action of all these forces will shape the trajectory of advances in technology, in the Auto ID industry and in the future supply chain.

One example: it is quite recent the announcement of the Alien Technology Corporation [1], that won an order from the Gillette Company for 500 million low-cost radio-frequency identification (RFID) tags, the first major commercial order for products incorporating the electronic product code, developed by researchers at the Auto-ID Center of the Massachusetts Institute of Technology. The same is doing clothier Benetton in adopting Philips' RFID for smart labels embedded into the labels of garments bearing the name of Benetton's core clothing brand, Sisley [2,3].

Labels are no more simple information elements but provide the vital links connecting information systems to the physical world of manufactured goods. In order to increase productivity, reduce costs and gain a competitive edge, logistics providers are seeking out technology-driven applications like real-time dispatch, pickup and delivery, loading of delivery trucks and dynamic route planning. The entire pipeline is now adopting information technology to provide on-line, real-time data for the movement of freight through their systems.

The logistics management is rightly considered the front line of business efficiency, where the physical goods in the distribution system (the packages) must be inseparable from the information system; in fact, it treats them as one and the same thing. Nicholas Negroponte [4] at MIT coined the phrase "translating atoms into bits" to describe this process. Jerome Schwarz, describing what it is done at Symbol Technologies, modifies that phrase, telling that they "translate atoms into bits in the form of photons" [5].

And we can assume this phrase as a remark, telling us that applied physics can give a strong benefit to logistic, developing new devices or improving the existing, with a deep impact in a specific field, that of security and anti-fraud systems. A crucial point of the supply chain, is to ensure the authenticity of documents in transactions (for instance, credit cards or shipping documents) but also to guarantee the labelled products with systems against tampering, counterfeiting and piracy, for consumer and producer safety [6]. Let us think, for instance, of cases involving pharmaceutical products. The World Health Organisation estimates that losses due to counterfeit drugs in the EU in 2000 represent approximately 5.8% of the market. It can have terrible consequences: fake meningitis vaccine is believed to have resulted in the deaths of 3000 people in Niger in 1996 [7]. Tests revealed that they contained no active ingredient.

Moreover, in the field of medical products, it is often unclear whether medical consumables have been adequately sterilised and whether they remain safe. Tools for single or limited reuse are not always monitored automatically for the number of uses and discarded in a timely fashion. Labels able to perform a monitoring are very important in all these cases. With the use of proper substances added to inks, labels can be easily transformed in sensors indicating temperature, humidity and many other excursions [8]. A popular application is the label on the bottle giving the right temperature of the beer. Probably, we will see very soon labels revealing bacteria, viruses, and poisons based on new semiconductor surface activation actually in development in research laboratories.

Before starting our discussion on the subject of label security, and in particular of methods to improve security based on applied physics, let us talk a little about the ID systems, which logistics providers actually use for goods identification in the supply chain. Then we shall see how these technologies can be extended to introduce anti-counterfeit elements.

**Labels with 1D, 2D and RF identification.**
Data are the basic building blocks of today's Information Economy and knowledge- based businesses. What's done with data, how they are processed, stored or otherwise manipulated, determines their value. For logistic workers, there are a number of key technology enablers for successful data management: e.g. one- and two-dimensional bar code (Fig.1), wireless data-voice communication, radio frequency identification (RFID) [9,10], wearable computers and other ergonomic designs. Moving into the world of high capacity symbologies, 2D systems can automatically read, store and process a huge amount of bytes of data. A 2D barcode can represent the actual data or graphics and it is then very useful in warehousing and transportation, where critical shipping information often becomes separated from the goods in travelling through the supply chain. If goods were processed without documentation, subsequent re-loading delays can be costly for carriers and consignees: with a 2D barcode, or Portable Data File (PDF), the shipping label, manifest, or bill of lading becomes a document more convenient to move, than data over paper. 2D codes greatly improve the logistics of the mail services for receiving and processing goods in transport at very low cost. 2D code is commonly used in post-offices (in USA, RFID labels are also used [11]).

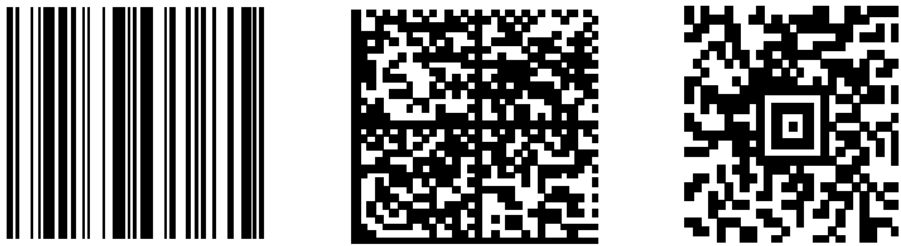

*Figure1.* *From left to right: 1D barcode, 2D Data Matrix, and 2D Aztec code.*

And in fact, the other Auto ID technology, which complements 2D systems and Portable Data Files with higher data capacity and read/write capability is the radio frequency identification (RFID) [12]. Delivering non-contact read and write capability, often with large data capacity, RFID tags span a wide range of cost/performance trade-offs. At the high-end, active tags (battery-energised) can communicate over ranges of hundreds of meters and store hundreds of kilobytes of information. Passive tags (see Fig.2) are instead cheap: they are simply done with a microchip and a metallized antenna coil on a thin polymeric sheet. Passive tags respond to a radio frequency signal.
Innovative technologies like wearable and wireless computers, and pen computers with integrated 1D and 2D scanning capability are transforming today's logistics. As the technology continues to accelerate, and performance capabilities are pushed forward, new technologies will not only shift toward a specific choice for automatic identification, and away from others, but will open up new applications, previously not conceived. Looking forward, Auto ID will evolve from a coexistence technology model to a complementary technology model, particularly useful in the logistics applications. For instance, 1D and 2D bar codes and RFID technologies can complement each other in a nested sequence, to provide a total automatic identification solution for generic transportation and logistics applications.

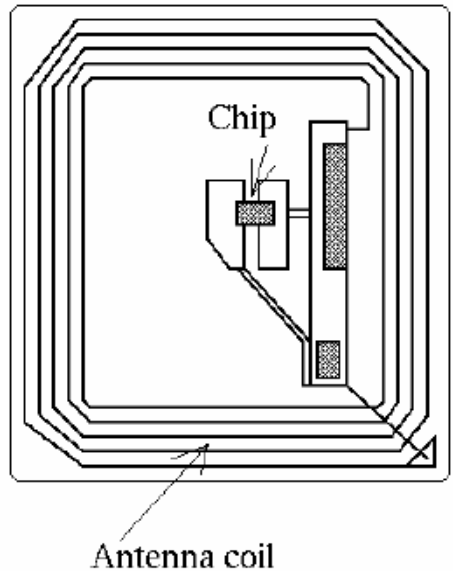

*Figure 2.* *Scheme of a passive RFID tag. On a label of small dimensions (few centimetres), the chip is connected with the antenna coil.*

Among the strongest trends in society, we find e-commerce and Internet. From banking and home office work to shopping and healthcare, the number and type of transactions conducted from home is growing. In fact, the future of security in all this kind of data transactions is not only demanded to the information science. Physics is involved too: on small local area networks, the quantum cryptography is working to ensure document transmission security [13]. Quantum mechanics is also proposed for secure no-clone smart cards [14].

Several methods can be used for labelling authentication and substances and instrumentation are under investigation to implement authentication. The term "substances" is usually covering any substance or composition of substances capable of being applied on the surface or into the bulk of any arbitrary object, locally or diffused, like ink, dye, glue, powder, film, wire, foil or as adhesive label itself. Among techniques actually used, we will discuss some of them, more connected to physical methods. Let us start with magnetic systems.

**Smart labels with microwires**

Smart labels are labels, which can give more information than the visible ones, usually printed on them. In the more simple form, the smart labels are bringing markers which can be detected by means of proper devices. One simple type of marker in the form of an electronic circuit comprises inductance and capacitance, elements that resonate at radio frequencies.

Another type of marker, a magnetic marker, comprises a strip of soft magnetic material that interacts with a ferromagnetic element made of a hard magnetic material that can be magnetized or demagnetized. The soft magnetic strip resonates and generates harmonics in the presence of a magnetic field having a certain frequency. This allows the marker to be identified. The hard ferromagnetic element can be magnetized or demagnetized, thereby deactivating or activating the marker. Among magnetic markers, we find the magneto-mechanical ones. They comprise a strip of magnetostrictive material and a strip of magnetic material of high coercivity. The magnetostrictive material mechanically resonates in presence of a magnetic field of a particular frequency. Using a receiver, sensitive to the magnetic field created by the mechanical resonating magnetostrictive material, we can detect the resonance. Modifying the magnetic bias of the magnetic material strip deactivates the marker.

Magnetic material are now strongly investigated to build low cost electromagnetic identification (EMID) tagging solutions. The global market for the conventional RF identification is already several billion dollars yearly and growing 20-30 % each year. These tags typically cost dollars to tens of dollars each. However, there are many inventions of EMID tags that cost, or are promised to cost, 50 cents or less. The point of view is that these tags will create even wide markets than those for conventional RFID.

Some of these smart labels are now based on microwire or tiny flecks, suitable for remote detection of tampering, remote counterfeit detection and logistic control of even the smallest jewellery, engineering parts, medical disposables and drugs, etc. Holotag in the UK and Remoso in the Netherlands [15,16] guess that the next generation will consist of arrays of specially prepared microwires, crossed like an asterisk or aligned like a barcode. Wire structures can have a smaller 'footprint', i.e. higher data density and lower cost, at least. Potentially, they can be as thin as all but the thinnest thin film chipless tags. 'The Internet of Things' envisaged at a price of only a few cents [17].

MXT Inc., start-up company in Montréal, Canada, is specialising in making magnetic wires [18]. Its mission statement is to be the world's leading supplier of low-cost, high performance electromagnetic markers for article surveillance, article location and identification. MXT is not an RFID company, in the sense of companies we mentioned before (Alien or Philips); it seeks to be a materials company, with the key strategy to give chipless smart label wires, providing the aforementioned benefits at only a few cents or tens of cents per tag assembled using an array of wires [18].

The company produces a soft magnetic wire of around 30 micrometers in diameter in lengths from 1 to 50 meters long. A further development is to extend this process to different materials and to develop their technology by varying size, length and material of the wire, to give different resonant signatures, designating particular lot to a particular vendor, in this way without carrying digital data. This would be used in brand authentication applications.

**Magnetic barcode on labels**

In the remote magnetic sensor identification, an equivalent of the barcode is created by inserting little magnetic elements opportunely spaced in a thread of a synthetic material (for instance PET) or on a sheet, and to remotely detect the position of these elements. The problem is remote detection: the more recent technique is based on the so-called Flying Null detection. FN tags can give each product an individual identity enabling covert brand protection, through authentication and tracing. The magnetic sensing technology can remotely determine the precise position in space of magnetic material, with many of the benefits of simple RFID but typically at a fraction of the cost. The size of these thin-film tags and their resistance to heat and pressure allow embedding into virtually any product and packaging during manufacturing.

The measuring method is build on the following principle. When two bar magnets are placed in close proximity with like-poles facing, we find a volume between the two bars where the fields from the two magnets cancel out. This region is called the 'null'. Let us introduce small amounts of very soft magnetic material between the magnets: outside the null region the material is saturated by the local magnetic field, however, in the null region, it is not saturated. If we introduce a time-varying magnetic field, the soft magnetic material in the null region will respond to the varying field and the response can be remotely detected with suitable antenna coils. Outside the null region, the soft magnetic material cannot respond because it is saturated. It is then possible to determine when there is a piece of material in the null and, scanning the null, locate the position of any piece of material in this region.

Having established the position of one element, it is possible to extend the concept by using many elements, producing a codable tag. Like an optical bar code, this can be linear or two-dimensional, with gaps between elements defining the data content of the tag. Small elements of soft magnetic material of less than 1 micron in thickness can be mounted on a thin layer of PET, typically 25 microns thick. The magnetic elements are a few millimetres long and a few millimetres wide. Actual dimensions are dependent upon the reading range required.

The company, named Flying Null, spinout from Scientific Generics Ltd. an international consulting and investment group, is producing the new tags for pharmaceutical security [19]. Since tags can be read through the aluminum of a blister pack, or a bottle cap, the packaging does not have to be opened to confirm authenticity of the contents. Flying Null announced that tags were applied as a thread, 23 microns thick, in between the plastic and the foil of blisters, in products of a global pharmaceutical manufacturer (Fig.3). It is believed this is the first time a tagging solution has been effectively embedded within blister packaging.

These new tags can be a cost-effective way for pharmaceutical manufacturers to combat counterfeiting of drugs and pharmaceuticals. Like RFID tags, FN tags can be read through cardboard and plastic; they can be read through the aluminum too, but can't be used with any magnetic material. These taggants just few microns thick can be embedded in the seals at point of manufacture and ensure that the contents have not been altered or tampered at any point in the supply chain [20].

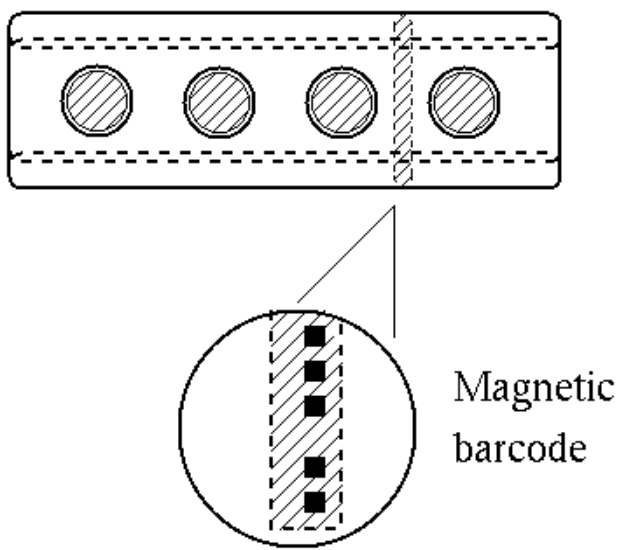

*Figure 3. Soft magnetic barcode of the Flying Null label inserted in a blister. The length of each magnetic element is about 1 mm.*

**Micrometer and submicrometer barcodes**
Originally developed by the 3M Corporation, the Microtaggant® brand Identification Particle is a microscopic barcode anti-counterfeit technology, highly versatile in its use and application [21]. A microtaggant particle is done by means of a chemically stable material consisting of several layers (typically eight or nine for those used with explosives) of a highly cross-linked polymer. Each layer has a different colour, and the colour sequence can be translated to a numeric series according to a colour codes. Ferromagnetic and/or fluorescent layers are incorporated for detection and retrieval. The multilayer slab is then broken in small pieces. The unique colour code sequence is produced for each customer or application, certified and registered as a product fingerprint. With just the basic coloured layer structure, it can produce over 37 million unique codes. The first application was the tracking of explosive materials [22].

The particles used for explosives are about the size of ground pepper. But different sizes can be made, down to 44 µm in mesh size. The material is mixed thoroughly with the explosive so it is part of the product. When a tagged explosive is detonated, the thermal effects of the blast will destroy most of the particles, but the population of taggants in a given quantity of explosive is high enough to ensure that some particles survive. These can be found with a magnet or with ultraviolet light, and the colour sequence can be read visually with a pocket microscope of at least 100x magnification (Fig.4).

The Swiss government passed a law in 1980 requiring explosives to contain a marker substance that allows reliable determination of origin after detonation. According to the Swiss Scientific Research Service (SRS) in Zurich, the law was prompted by the world-wide spread of terrorism in the 1970s, and its primary purpose is to assist the police in investigations of criminal offences involving explosions. 3M microtaggants are used also like an anti-counterfeit substance in pharmaceutical product labels.

Recently multimetal microrods intrinsically encoded with submicrometer stripes have been synthesised, creating barcodes of the length of few microns (see Figure 5). This novel method of encoding information could be used for the simultaneous detection of numerous biological analytes, the main purpose of the research team. The Pennsylvania State University and SurroMed, a biotechnology company in Mountain View, California, carried out the work. Varying the sequence of metal electrodeposition [23], it is possible to have different bar codes.

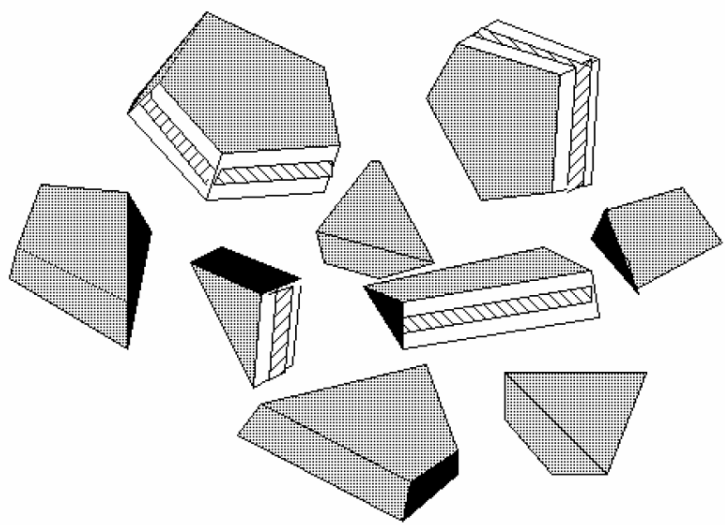

*Figure 4. Layers of different colours are used to obtain 3M microtaggants. These particles can be easily identified in a mixture with other particles with a microscope.*

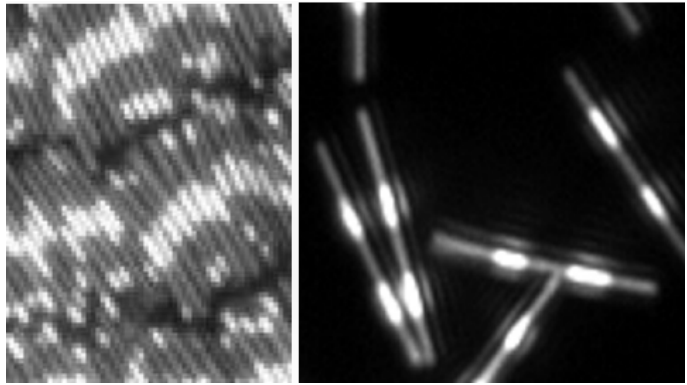

*Figure 5. An example of barcode assembly (on the left) and Au/Ag barcodes (on the right). Each of these particles is about 4.5 microns long. Alternating gold and silver stripes create the "barcode" pattern on the particles. When viewed in blue light under a microscope, silver is much more reflective than gold, making different-patterned particles easy to identify.*

The barcoded microrods, which comprise cylindrical segments of up to five different metals, are prepared by sequential electrochemical reduction of metal ions into the pores of an alumina membrane template with a silver film as its base. The length of stripes in the code (between 50 nm and 5 µm) is also varied by changing the amount of electrical charge passed through each metal ion solution. The barcoded particles are released by dissolving the template and the silver film in nitric acid and sodium hydroxide solution, respectively (Fig.6). The team uses conventional light microscopy to identify the bar codes. The method relies on the differences in optical reflectivity between adjacent metal stripes.

As pointed out by the researchers, microrods with different bar codes can, in principle, be coated with different reagents that capture a wide variety of biological molecules such as DNA sequences and proteins. A mixture of barcoded microrods could then be added to a solution of blood plasma, for example, and a standard technique such as fluorescence imaging used to detect and quantify analytes that bind to the barcoded microrods. The identity of an analyte would be determined by the bar code reflectivity readout (Fig.7).

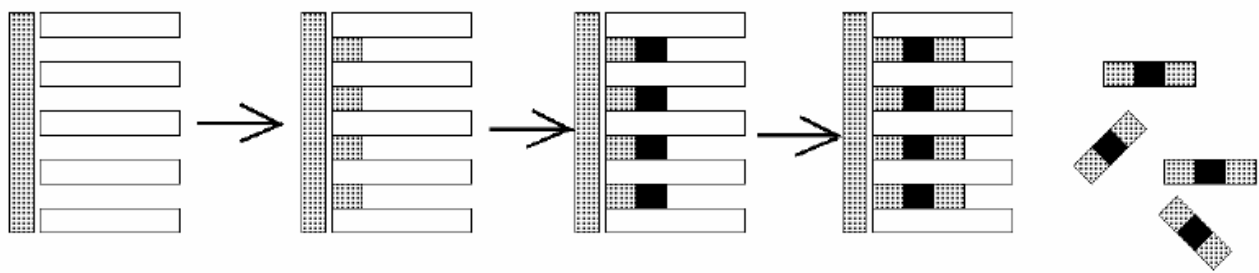

*Figure 6. Procedure for preparing barcode nanorods (from left to right). Into the pores of a metallized template, a sequential electrochemical reduction of metal ions is made. Dissolving the template, a nanobarcode suspension is obtained.*

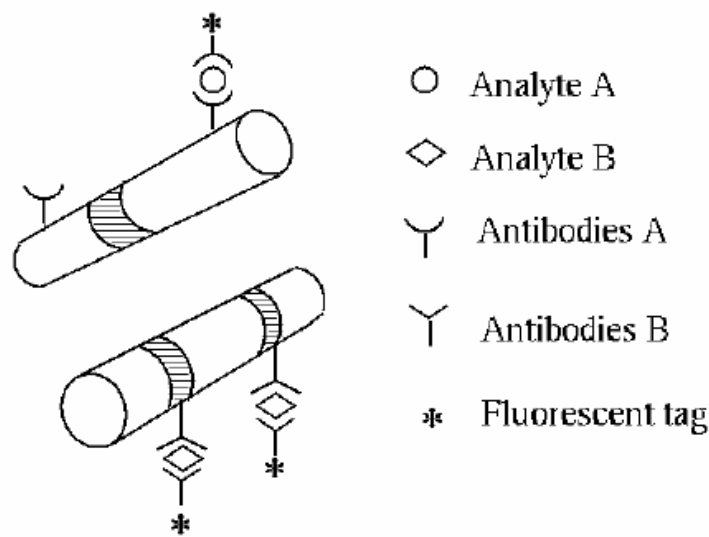

*Figure 7.* *Barcodes with different reagents on them. The code is used to distinguish the reagents.*

Of course, the nanobarcode-particles technology holds great promise in non-biological applications, for robust, identifiable nanoscale tagging of small items for authentication or tracking. Cunin et al. [24] produced microbarcodes in porous silicon with a well-ordered arrays of micropores using photolithographic structuring before the electrochemical formation process. They chose to identify the different silicon beads by their individual interference colour, obtaining in some sense a spectral barcode: it is in fact the interference analogous of the 3M microtaggant, since colours are based on light interference.

Another technique for tagging labels can be the microencapsulation of dyes or other substances: this technique can be useful for textiles [25]. Let us now pass to discuss tagging methods that are using the atomic scale, that is methods based on the electron paramagnetic (EPR), nuclear magnetic (NMR) or nuclear quadrupolar (NQR) resonance techniques.

**EPR and NMR in external magnetic fields**

A first patent which claimed the application of electron paramagnetic resonance (EPR) to the problems of authentication or identifying papers was patent US4376264 [26]. It disclose us the use of substances having EPR characteristics detected in high field EPR in the microwave band. Other patens claim the use of EPR in radiofrequency band or the solution of authentication problems with the nuclear magnetic resonance (NMR), ferromagnetic resonance, ferrimagnetic resonance and so on [27]. All the patents are based on a variety of magnetic resonance phenomena. These phenomena are associated with nuclear, electronic, atomic or molecular magnetic dipoles moments acting in individually or cooperatively way in the presence of external magnetic fields to give the resonance. The magnetic resonance is exhibited when the dipole moments precessing in the magnetic fields absorb and re-radiate microwaves or radio-frequency electromagnetic radiation at or very close to the precession frequency. The serious disadvantage of applying the methods is the necessity of an external static magnetic field. This magnetic field can be supplied either as a large field over the entire interrogation volume, or by a small permanent magnet close to the resonant material and carried around with it. It is a rather complicated design for labels and reading devices: moreover, there is the risk of wiping out data contained in magnetic media.

**NMR zero-field and NQR labels**

The magnetic ordering of a solid with a spontaneous magnetization is clearly revealed by the resulting macroscopic field. However the magnetic order in antiferroelectric solids yields no macroscopic field, or in ferromagnetic it is often masked by a domain structure. Nuclear magnetic resonance offers a way to probe the microscopic spin structure, since the ionic nuclei fell the dipolar magnetic field of nearly electrons. Nuclear magnetic resonance in magnetically ordered solids can therefore be observed even in the absence of applied field, the field at the nucleus (and hence the resonance frequency) being entirely due to the ordered moments.

The previously described zero-field NMR, is commonly confused with the Nuclear Quadrupole Resonance. NQR is a valuable way of observing nuclei with spins > ½ (examples of such nuclei are deuterium, $^{14}$N, $^{37}$Cl, $^{63}$Cu, $^{67}$Zn and $^{127}$I). The nuclei have a nuclear dipole moment, which makes them NMR-active; in addition they also possess a nuclear electric quadrupole moment that interacts with the gradient of the local electric field, at the nucleus site, giving an electric quadrupole coupling. The size of this coupling depends on the value of nuclear quadrupole moment and on the field gradient at the nucleus. Often the electric quadrupole coupling is larger than the conventional NMR frequency, even for the largest magnets currently available. It follows that it is not necessary to have a magnet for NMR; the magnetic dipole resonance in magnetically order solids or the electric quadrupole couplings will give us an energy splitting useful enough to investigate the material. The nuclear quadrupole resonance (NQR) is used as a bulk inspection technology for detecting crystalline explosive solids containing $^{14}$N nuclei such as RDX, TNT, and nitrates [28]. Since NQR takes advantage of the natural crystalline electric field gradient, the result depends on the chemical structure of the material and then it will be different for RDX, for TNT, etc.. The electrical field gradients align electric quadrupole moments, due to the non-spherical nuclear charge distribution of the $^{14}$N nuclei. When a low-intensity RF signal is applied at certain frequencies, usually in the range 0.5 to 6 MHz, the alignment of $^{14}$N nuclei is altered. As the RF is removed, nuclei precess to their original state (actually a transition between the energy states resulting from the previously described interaction), producing a characteristic radio signal [29,30].

For what concerns the possibility to use zero-field resonance in labels, Micro Tag TEMED is commercializing an anti-counterfeiting system based on zero-field NMR and NQR labels [31,32].

In fact, we can imagine to use other methods to investigate a product; for instance, organic analytical detection methods, such as infrared spectroscopy, gas and liquid chromatography, and mass spectrometry, to separate and identify the mixture components. Moreover, the biological investigation of DNA can give us the geographical origin of the product [33]. Inorganic methods can also be useful tools, such as ion chromatography, atomic absorption, and inductively coupled plasma spectrometry to measure inorganic components of mixtures. Among inorganic methods we must consider image analysis too, with optical and scanning electron microscopy helped by digital processing of data. Microscopy is the first step to detect physical evidence of tampering or counterfeiting. All these techniques are very interesting research field of the forensic science, but we will not consider in this paper.

Let us instead discuss how optical methods can be applied for label authentication. A wide variety of optical security features are existing, ranging from watermarks and intaglio printing, via tilt images, fluorescence and digital anti-copy devices, to optically variable diffraction and interference based security devices, named "Diffractive Optical Image Devices".

**Diffractive Optical Image Devices.**
Holography was discovered in 1947 by Dennis Gabor, who received the Nobel Prize in Physics in 1972. This technique is the only visual recording/playback process that can record the three-dimensional world on a two-dimensional recording medium and playback the original object or scene, to the unaided eyes, as a three dimensional image. The image demonstrates complete parallax and depth-of-field, floating in space behind the recording medium [34]. Holograms belong to a class of images known as Diffractive Optical Variable Image Devices (DOVIDs) and are rapidly becoming the quintessential method of protection against counterfeiting, method based on diffractive optics.

Highly valued for security, DOVIDs can only be produced using expensive, specialized and technologically advanced equipment. They cannot be replicated by colour copiers, computer scanning equipment, or by standard printing techniques (analog or digital) since they are based on diffraction of light, whereas printing reflects it. Given that counterfeiters inherently select the path of least resistance, encountering a security hologram will probably cause them to abandon their efforts and move on to easier targets. As a result, the use of hologram security devices virtually guarantees product authenticity.

Security holograms have become well-established in many industries and are commonly found on a host of products and packaging, including compact discs, computer software, cosmetics, watches, and sporting goods, no more only on Visa® and Master Card® credit cards.

Several type of holograms are used. We find holograms embossed for mass production in a plastic film or metallised on photopolymeric photosensitive films, usually on a polyester film substrates, which produce high-resolution, relatively bright and easily viewable holograms, with dichromated gelatin (DCG) or silver halide as photosensitive medium.

Usually, holograms are played in white light. In transmission, a hologram is illuminated from the side opposite to the viewer, so that the light is transmitted through the hologram. This typically creates a rainbow -colour image, although special colour and true-colour effects can be created. In reflection, the hologram is illuminated from the viewer side, typically creating a single-colour hologram, although pseudo-colour and true-colour effects can be created. Embossed holograms are all white light transmission holograms which are metallized or otherwise treated so that the transmitted light reflects back to the viewer, who therefore sees the hologram in the reflection mode.

There is a direct relationship between the type of hologram and the lighting required for viewing it. Holograms comprising only surface images (2D) can be adequately seen in ambient light, but the more depth and complexity there is in the hologram image, the smaller and brighter the illuminating light source should be to do the hologram justice. A single incandescent bulb, preferably clear, is adequate for most smaller 3D holograms. Some covert images may require illumination by a light source of a specified wavelength, such as a laser, laser diode or LED.

Let us note that DOVID is a generic term used in the security printing and authentication field. Optically Variable Devices (OVD) is a term used to describe several types of authentication, anti-copy component (including inks, thin films and foils); manufacturers introduced proprietary techniques and device names, such as Kinegram® and Exelgram® [35,36]. DOVIDs include all types of hologram and other diffractive optical foils or film products.

Micromanifacturing allows to produce Exelgram®, an optical security technology developed by CSIRO, Australia for the protection of banknotes and high security documents [37]. It is typically mass produced in hot stamping foil: the surface has a relief structure which, when illuminated by a light source, generates one or more diffraction images that are observable from particular ranges of viewing angles around the device. At least, part of the surface relief structure is arranged in a series of tracks, including tracks of variable groove angle and spatial frequency, obtaining then the optical variable effect.

Another specialised Exelgram® feature is the two-channel right angle optical effect. In this effect, two distinct images are encoded into the Exelgram® and replayed to the observer by the simple expedient of rotating the device by ninety degrees in its own plane.

The optical variable labels can take advantages from the micrographics techniques too. The image is then designed to incorporate very small scale graphic elements consisting of combinations of alpha-numeric characters and other graphic elements such as logos and line drawings, in the range of feature sizes from 1 to 30 microns. These elements act to diffusely scatter the incoming light. Particular advantages of such hybrid diffractive/micrographic OVD structures include: easy microscopic forensic authentication of the smallest piece of embossed substrate; extremely high security against attempted holographic copying; and, reduced metallic appearance of the OVD foil due to diffuse scattering from the micrographic elements. As discussed in Ref.[38], holograms are more easy to copy that is widely believed though advanced techniques, such as micrographic technique make duplication far more difficult. Zero-order diffraction microgratings can be a more secure alternative.

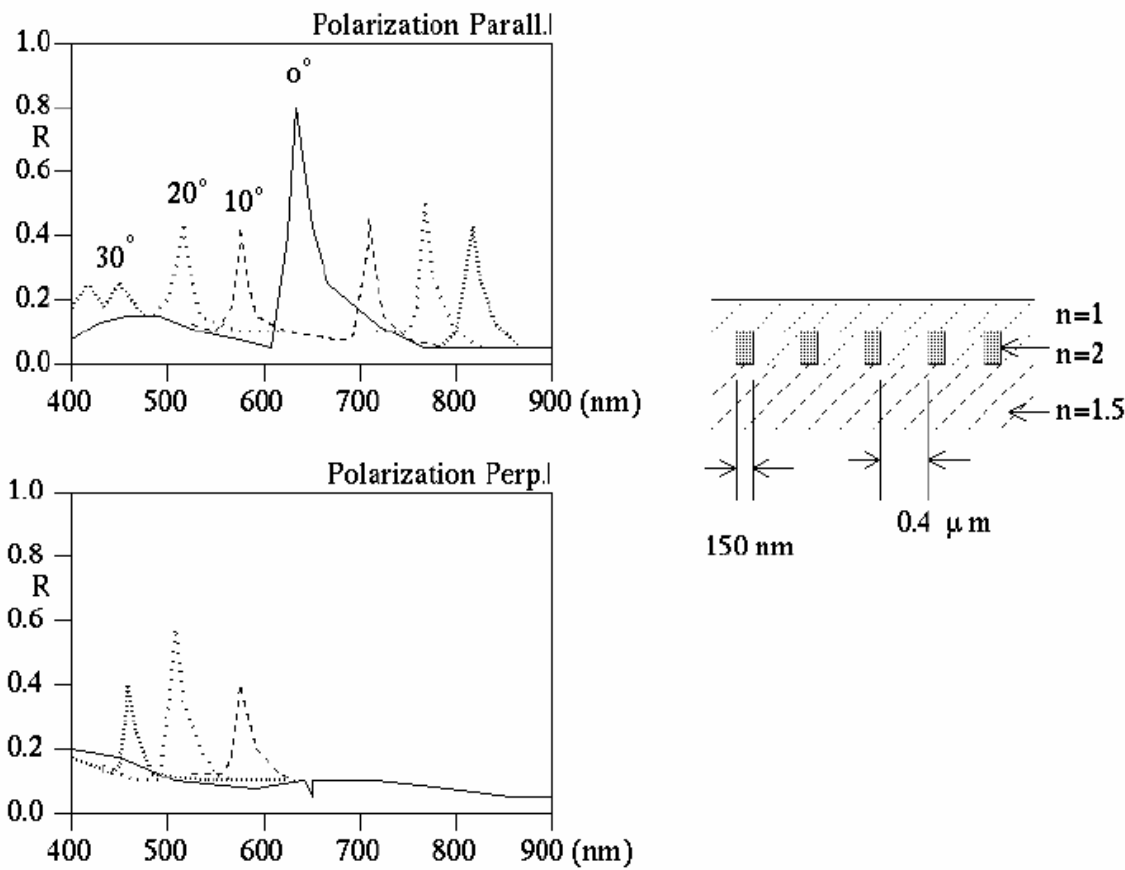

*Figure 8*. *Reflection of the zero-order diffraction grating as a function of wavelengths.*

Zero-order diffraction microgratings are structures with a characteristic length, in the hundreds of nanometers, less then that of visible light, and only specular reflection will be present. The structure has a very pronounced reflectance peak in the red portion of the spectrum at normal incidence for light polarized parallel to the grating lines. The peak splits and shifts with the angle of incidence, leaving the visible range at 30°. For light polarized perpendicular to the grating lines, the peak is in the green portion of the spectrum: rotating the structure about its own axis will result in a colour change from red to green. Fig.8 shows data of a microstruture reported in Ref.[39]. Microgratings are made of plastics, hence can be embedded within a security device during manifacturing process and are very well suited for machine verification. Zero-order microgratings offer great resistance to duplication. Moreover, the manufacturing capability necessary to make the structure is rather expensive.

Let us now discuss a very cheap and easy to implement optical technique for label security.

**Intrinsic check of authenticity**

Intrinsic systems, based on optical detection of random features in combination with digital signatures based on public key codes in order to recognize counterfeit objects, can be easily developed for protecting banknotes and labels. Without applying expensive production techniques objects are protected against counterfeiting. The verification is done off-line by optical means without a central authority [40]. The method is based on the fact that microscopic random features like the fibre structure of paper or air bubbles in plastic are indeed very hard if not impossible to copy in a cost-effective way. The key idea is that every object to be secured against counterfeiting has its own random security feature. These features do not have to be produced because they are already present on the objects to be secured.

Recently [41], it was proposed a security label system based on the same principle, to search the presence one or more distinctive signs casually applied onto a support. As in Ref.39, a detection with a macro-imaging of the marking is done, a macro-image including data concerning these distinctive signs as well as data concerning the structure of the support onto which signs are applied. The fundamental concept is to couple such intrinsic elements of marking with the design of the basic structure of supporting media. The combined image, formed by the structure of the support and the distinctive marking signs is simply detected by using a photomacrography technique. The required magnifying ratio varies according to the type of used support (except in special cases, 5 and 10 magnifications). The photomacrography has to be taken in the most suitable lighting conditions for the marking, to show up as far as possible the structure of the support on which the marking is applied.

In the case of particularly homogeneous supports, not apt to show up an intrinsic differentiated structure, they can be subjected to a previous surface treatment allowing to place in evidence a inherent structure, for example through a mechanical machining or through a selective etching or with an additional layer having a non-homogeneous structure applied thereon. Once the reference is made, it is then possible to check the combination between markers and reference (an example for a label with a reference under the marks is shown in Fig.9).

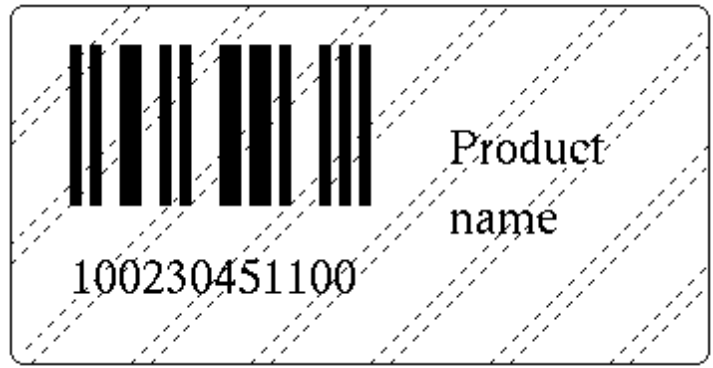

*Figure 9. An example of a secure label, with a reference frame under the printed elements.*

In this paper we have discussed few examples of the new smart labels, used as antifraud, anti-counterfeiting systems. They are case studies in which it is possible to see immediately, how applied physics can deeply modify and improve security systems. Many other systems based for instance on nanostructures can have interesting applications in the field of security. Probably we shall see in the next future many new companies, working in the security area, spinout form physics research laboraotries.

**Acknowledgement**
The author is indebted with Ermanno Rondi (Incas Group, Vigliano Biellese) for useful discussions.

This paper is a version with upgraded references of the manuscript published in Recent Research Developments in Applied Physics, 2003, Vol. 6, pp.39-54, ISBN: 81-7895-085-5